\documentclass[useAMS,usenatbib,referee]{mn2e}

\usepackage{graphicx}
\usepackage{longtable}
\usepackage{times}
\title[A potential new method for temperature determination in LMS]{A potential new method for determining the temperature 
of cool stars}
\author[S. Viti et al.]{S. Viti$^1$, H. R. A. Jones$^2$, M. J. Richter$^3$, R. J. Barber$^1$, J. Tennyson$^1$, J. H. Lacy$^4$ \\
$^1$Dept. Physics and Astronomy, University College London, Gower Street, London, WC1E 6BT, UK \\ 
$^2$ Centre for Astrophysics Research, University of Hertfordshire, College Lane, Hatfield, Hertfordshire AL10 9AB, UK \\ 
$^3$ Physics Department, UC Davis, One Shields Ave, Davis, CA 95616, USA \\
$^4$ Department of Astronomy, University of Texas at Austin, 1 University Station, Austin, TX 78712, USA \\ }

\date{Released 2002 Xxxxx XX}

\pagerange{\pageref{firstpage}--\pageref{lastpage}} \pubyear{2002}

\def\LaTeX{L\kern-.36em\raise.3ex\hbox{a}\kern-.15em
    T\kern-.1667em\lower.7ex\hbox{E}\kern-.125emX}

\begin{document}
\maketitle

\begin{abstract}
We present high resolution (R = 90,000) mid-infrared spectra of M
dwarfs. The mid infrared region of the spectra of cool low mass stars 
contain pure rotational water vapour
transitions that may provide us with a new methodology in the
determination of the effective temperatures for low mass
stars. We 
identify and assign water transitions in these spectra and 
determine
how sensitive each pure rotational water transition is to small (25~K) changes in
effective temperature. We find that, of the 36 confirmed and assigned pure rotational water
transitions, at least 10 should be sensitive enough to be used
as temperature indicators.

\end{abstract}

\section{Introduction}

Low mass stars (LMS) constitute $\sim$80\% of our stellar
neighbourhood.  They provide a probe of our understanding of main
sequence stellar evolution and are key in determining the boundary
between stellar and sub stellar objects. However, their spectra are
extremely rich in structure and their opacity is made up of many
molecular and atomic absorbers, each with hundred of thousands to
millions of spectral lines. This means that colours are not easily interpretable as
diagnostic of their properties.  Among the fundamental
properties of LMS, of particular importance is their effective
temperature, T$_{eff}$: the latter has frequently been investigated
but it is still not well determined. In particular there is not yet a
tight correlation between spectral type and effective temperature
as there has been a long standing discrepancy between empirical
effective temperatures and those derived by synthetic spectra
($\sim$ 125~K as estimated by Kirkpatrick et al. 1993).  This is primarily due to the lack of a
complete inclusion of molecular and atomic opacities in the models, in
particular in the near infrared (NIR).  Modelling the atmosphere of LMS in the NIR
is not a trivial task: water vapour dominates this part of the
spectrum and at the effective temperature applicable to cool star
atmospheres ($<$4000 K) water can access energies as high as 45000
cm$^{-1}$ before it dissociates; to reproduce high temperature water
spectra is very challenging because of the complexity of the vibrational and rotational motion of
asymmetric triatomic molecules.  Leggett et al. (2000, 2001) used a
synthetic grid which includes the water linelist calculated by
Partridge \& Schwenke (1997) to determine fundamental parameters for a
large sample of LMS. They conclude that problems remain with the match
of the observed water bands.  Allard et al. (2000) also find that the
models including this water linelist reproduce well late-type dwarfs
but fail to reproduce hotter dwarfs. The main sources of error in
present water opacity databases are the incorrect high temperature
transitions (Jones et al. 2002, 2005).

\par 
Recently a new water linelist, BT2, has been produced (Barber et
al. 2006).  The BT2 linelist is by far the most complete linelist
available, reaching higher energies than previously published. A
comparison between a previous release of the BT2 linelist (BT1) and
the PS linelist showed that at higher temperatures (T $>$ 2000K) the
PS linelist is missing around 25 per cent of the water vapour opacity
(Jones et al.  2005). While the BT2 increases the accuracy of the
fitting in the NIR region, its cut off at energies above 30000 cm$^{-1}$ still means that none
of the energy levels above 20000 cm$^{-1}$ are capable of being excited
by 1 $\mu$m photons (Barber et al. 2006).  \par Although the strongest
water vapour opacity is at shorter wavelengths, analysis of sunspots
(Wallace et al. 1995) shows that H$_2$O also accounts for the majority of
lines in the mid infrared (MIR) region between 8 and 21 $\mu$m. Polyansky et al
(1997a,b) successfully assigned many of the water transitions in the
10-13 micron spectra of sunspots with further recent assignments being made in conjunction with the analysis of oxy-acetylene emission spectra (Coheur et al. 2005; Zobov et al. 2006). In this spectral region the water
transitions are largely pure rotational, in contrast to the
vibration-rotation transitions which dominate at shorter
wavelengths. The use of pure rotational transitions greatly simplifies
the spectral analysis as estimates of the transition strengths are
much more straightforward. To a good approximation the strength of
individual rotational transitions can be estimated from a simple
algebraic formula times a Boltzmann factor. Furthermore, the presence
of pure rotational transitions within different vibrational states,
which is characteristic of the sunspot spectrum, yields a large
dynamic range for temperature analysis. These advantages potentially make the
MIR region ideal for the determination of the effective temperature
scale for LMS.
\par Here we propose a new methodology for establishing the 
effective temperatures of cool stars, in particular LMS, based on high resolution MIR
spectroscopy. Taking advantage of the relatively easy to interpret
MIR spectrum in cool stars has been done before: 
Ryde et al. (2006)
used high resolution observations of water vapour in super-giants 
to show that
classical photosphere models can not fit the observed spectra and that
synthetic spectra based on cooler photospheric temperature structure
are needed. This supports the idea that individual rotational
water lines are highly sensitive to temperature variations in this
region.  In this paper we present the {\it first} high resolution (90,000) 
MIR
observational data for two M dwarfs, together with one M giant. Our observations
are described in Section 2. In Section 3 we present the data, the
identification of the water lines and a
theoretical sensitivity study using the BT2 linelist. In Section 4
we briefly conclude.

\section{Observations}

The data were taken in November 2006 using the Texas
Echelon-cross-Echelle spectrograph (TEXES, Lacy et al. 2002) at the
Gemini North 8m telescope. The observations were made on the nights of
Nov 20, 23, and 27 for a total of $\sim$13 hours.  We observed three
objects (BS587, GJ411 and GJ273 - see Table 1) in the TEXES high spectral resolution mode in two grating
positions: 11.2 and 12.5 $\mu$m (800 and 892.857 cm$^{-1}$). The two grating positions were chosen
on the basis 
of a  relative lack of narrow telluric features and abundance of
temperature sensitive water features based on an early version of our ab
initio water line list, as well as of a careful analysis of the sunspot (T$\sim$3300K) 10-13
$\mu$m spectrum.  The spectral coverage was 0.5\%\ at 11.2 $\mu$m and
0.75\%\ at 12.5 $\mu$m.  At 12.5 $\mu$m, there are slight gaps in the
spectral coverage because the angular width of the spectral orders is
larger than the detector.  The spectral resolution, as determined by
Gaussian fits to telluric atmospheric lines in the 12.5 $\mu$m
setting, has a FWHM equivalent to R=90,000.  We expect the same
spectral resolution at 11.2 $\mu$m.  The slit width was
0.5$^{\prime\prime}$ for all observations and the length was roughly
4$^{\prime\prime}$ and 2.5$^{\prime\prime}$ at 11.2 $\mu$m and 12.5
$\mu$m, respectively.  For all observations, we nodded the source
along the slit roughly every 10 seconds.

Of our targets, GJ411 and BS587 were sufficiently bright to fine
tune pointing based on the signal recorded from every nod pair.
GJ273 was weaker and required accumulation of several nod pairs to
evaluate pointing adjustments with confidence.  On Nov 23, clouds were
occasionally present based on elevated background signal through
TEXES.  We discarded the data that were most severely effected.

Coordinates and properties of the three objects are listed
in Table 1. The two dwarfs were chosen among a sample of well studied dwarfs:
note that for both GJ411 and GJ273, the quoted temperatures
have an uncertainties of at least 100K (cf Kirkpatrick et al 1993 and Tsuji et al. 1996). 
The quoted radial velocities are from the literature (Marcy et al. 1987) and they  
match the sunspot features within 0.5 km/s.

The data were reduced according to standard procedures as described in
Lacy et al (2002).  The TEXES pipeline provides for differencing of
individual nod pairs, correcting spikes and cosmic rays, establishing
a wavelength scale based on a user-identified telluric line,
correcting distortions in the instrument, extracting 1D spectra using
optimal weighting along the slit length, and combining spectra from
the same night weighted by the square of their signal-to-noise ratio.
The pipeline also performs flat-fielding and a first order correction
for telluric spectral features using the difference between an ambient
temperature blackbody and sky emission. Further flat-field and telluric
correction comes from division by a featureless point source such as
an asteroid or hot star.  Data from separate nights were aligned
spectrally and combined according to the square of their
signal-to-noise ratio using custom IDL procedures. The
dispersion calculations are accurate to 0.6 km/s or better.

\begin{table*}
    \caption{Properties of objects. }
    \begin{tabular}{|ccccccc|cc|}
Object & RA & Dec & Sp. & d (pc) & v$_{rad}$ (km/s) & M$_{bol}$ & T$_{eff}$(K) & log g \\
\hline
GJ411$^1$ & 11 03 20.2 &  +35 58 11.5&  M2V & 2.52 & -84.74 & 8.88 & 3510 & 5 \\
BS587$^2$ & 02 00 26.8 & -08 31 25.9 & M4.6III & 182 & +12.70  &   & 3425 & -1\\
GJ273$^3$ & 07:27:24.5 &  +05:13:32.5 & M3.5V & 3.76 & +18.10  & 9.56 & 3150 & 5  \\
\hline
\multicolumn{9}{l}{$^1$ Effective temperature taken from Tsuji et al. (1996); surface gravity from Jones et al. (1996); M$_{bol}$ from Cushing et al. (2005).} \\
\multicolumn{9}{l}{$^2$ Effective temperature taken from van Belle et al., (1999); surface gravity from Fluks et al., (1994). } \\
\multicolumn{9}{l}{$^3$ Effective temperature taken from Tsuji et al. (1996); surface gravity from Jones et al. (1996); M$_{bol}$ from Delfosse et al. (1998).} \\ 
\end{tabular}
\end{table*}

\section{Analysis}

Figures 1 and 2 show the recorded spectra for our sample for the two
wavelength regions. The gaps in the spectra coincide with the removals of
bad pixels and of telluric lines. The signal to noise varies from star to star
($\sim$ 100 for BS587, 40 for GJ411 and 5 for GJ273). We used the
bright star BS587 as a template for the identification of the water
lines in the spectra due to its high S/N. In total we identified 71 lines in the observed spectra which could potentially
be water (see Table 2);
however, only 52 of them are confirmed by comparison 
with the BT2 linelist (see Table 2 and Figures 3 and 4), some of these (24, 25, 29, 38, 44, 53, 62 and 66 in Table 2) are weak according to theory. Note that in order for a feature to be confirmed and assigned as water we required its intensity to be $>$ 10$^{-23}$ cm/molecule and its
frequency to be within 0.001 cm$^{-1}$. 40 of the 52 lines have been independently
identified and assigned in the Sunspot spectra (Zobov et al. 2006). 
Of the 19 remaining lines not identified in BT2, 13 are
also present in the sunspot spectra (Wallace et al. 1995) and 4 of these 13 have been in fact assigned to water (Zobov et al. 2006 and private communications). 
Many of the observed features are blends
of multiple lines (often 2, the para:ortho pair lines): for these cases we
assigned the band to the strongest component (ortho in the case of pairs of lines).

\par Figures 3 and 4 show examples
of our line assignment with the observed stellar spectrum on the top
panel overplotted to the sunspot spectrum (offset from the BS587 spectrum by +0.05), and with a synthetic water
linelist computed at 3500 K on the bottom panel. Note that the choice
of temperature for the synthetic water spectrum is
$not$ a best fit but was chosen to represent a bright cool stellar
atmosphere and is close to that of the sunspot spectrum. In fact, a synthetic spectrum at
3000K may be more representative of a 3500K giant since the gas temperatures of the layers where most flux is emitted are typically 500K less than the effective temperature of the star (Allard et al. 1997).
We have compared a 3000K synthetic water
linelist with BS587 and find that the same number of water lines are identified.

\par Some of the lines identified as water transitions in the observed spectra 
are not from experimentally-known levels. For some of these cases the quantum
numbers have not been fully determined and those given in Table 2 are the rigorous ones only 
from the ab initio data. We have singled
out these transitions in italic in Table 2. Finally, it is worth
noting that some of the identified lines are in fact ro-vibrational:
we have excluded these transitions from our sensitivity analysis (see
next section and Table 3).
\setcounter{table}{1}
\begin{table*}
\centering
    \caption{Transitions identified in the observed spectra of BS587. The confirmed 
water transitions are also given with energy levels, and   
quantum assignments when known experimentally. Transitions in italic are
experimentally unknown transitions. The 10th column (Sunspot) 
states whether the confirmed water transition
has been independently identified in sunspots (Y) or not (N).
The last column is the temperature of the maximum intensity gradient 
within the 1500 to
4000 K limits of our chosen model grid.
(see Section 4).}
    \begin{tabular}{|ccccccccccc|}
\hline
 & Frequency (cm$^{-1}$) & $\nu_1 \nu_2 \nu_3$ (u) & J (u) & K$_a$ K$_c$ (u) & $\nu_1 \nu_2 \nu_3$ (l) & J (l) & K$_a$ K$_c$ (l) & E(l) & Sunspot & T$_{max}$ (K) \\
\hline
1 & 797.130$^1$ &  100   & 18   & 7 12 & 100 & 17 & 4 13  & 7597.8219 &Y  &     1500 \\
2 & 797.188 & 000 & 24 & 11 14 & 000 & 23 & 10 13 & 7927.6588 &Y & 1500 \\ 
3 & 797.272 &     &    &       &     &    &       &           &Y  &      \\
4 & 797.301 &     &    &       &     &    &       &           &Y  &      \\
5 & 797.424 & 011 & 21 & 20 2 & 011 & 20 & 19 1 & 14390.7454 &Y & 3225 \\ 
6 & 797.551 & 010 & 20 & 17 4 & 010 & 19 & 16 3 & 9559.7822 &Y & 1500 \\ 
7 & 797.693 &     &    &       &     &    &       &           &Y  &      \\
8 & 797.966 & 020 & 20 & 13 8 & 010 & 19 & 18 1 & 10174.6070& Y  &1500\\ 
9 & 798.122 & 000 & 23 & 12 12 & 000 & 22 & 11 11 & 7656.5656&Y  &1500\\ 
\it{10} &\it{798.239} &  & \it{23} &  &  & \it{22} &  & \it{12650.7318} &  N & \it{2925} \\
\it{11} & \it{799.227} &  & \it{22} &  &  & \it{21} &  & \it{18066.4172} & N &  \it{3875}\\
12 & 799.333$^1$ & 011    & 21   & 19 3  &  011& 20 &  18 2 &  14106.5900  & Y  &3175  \\ 
\it{13} & \it{799.359} &  & \it{21} &  &  & \it{20} &  & \it{14106.5958} & N & \it{3175} \\ 
\it{14} & \it{799.372} &  & \it{22} &  &  & \it{21} &  & \it{12732.1876} & Y & \it{2925} \\ 
\it{15} & \it{799.493} &  & \it{25} &  &  & \it{24} &  & \it{15580.8988} & N & \it{3425} \\
\it{16}& \it{799.614} &  & \it{21} &  &  & \it{20} &  & \it{14671.8932} & Y & \it{3275} \\
\it{17} & \it{799.790} &  & \it{20} &  &  & \it{19} &  & \it{17771.2525}&N &\it{3800}\\ 
18 & 799.919 & 011 & 23 & 12 12 & 011 & 22 & 11 11 & 13031.5180 &Y &2975 \\
19 & 799.968$^1$ &010  & 13 & 7 7   & 010 & 12 & 4 8   &  3843.4105& Y & 1500 \\
20 & 800.105 &     &    &       &     &    &       &           & N &      \\ 
21 & 800.169 & 100 & 22 & 16 7 & 100 & 21 & 15 6 & 11774.0830& Y  &2750\\ 
22 & 800.229 &     &    &       &     &    &       &           & Y &      \\
23 & 800.280 & 000 & 24 & 9 15 & 000 & 23 & 8 16 & 7459.6821&Y&  1500\\ 
24 & 800.321 & 020 & 18 & 11 8 & 100 & 17 & 10 7 & 8608.2236 &Y& 1500\\ 
25 & 800.427 & 100 & 25 & 11 15 & 100 & 24 & 10 14 & 11962.4012 & Y& 2800 \\
26 & 800.435 &     &    &       &     &    &       &            & N&      \\
27 & 800.476 &     &    &       &     &    &       &           & Y &      \\
28 & 800.552 & 100 & 24 & 12 13 & 100 & 23 & 11 12& 11674.6517 &Y& 2750\\ 
29 & 800.611 & 020 & 34 & 2 33 & 020 & 34 & 1 34 & 14127.8724 &N& 3175\\ 
30 & 801.132 & 001 & 22 & 17 6 & 001 & 21 & 16 5 & 12109.9259 &Y& 2825\\ 
31 & 801.159  & 001 & 22 & 22 1 & 001 & 21 & 21 0  & 13486.1810  & Y &3075\\ 
32 & 801.219  & 000 & 21 & 17 4  & 000 & 20 & 16 5  & 8100.2905 &Y  &1500\\ 
33 & 801.252 & 000 & 24 & 11 13  & 000 & 23 & 10 14  & 7924.4497 &Y  &1500\\
34 & 801.360 & 000 & 21 & 21 0 & 000 & 20 & 20 1 & 9257.4587&Y  &1500\\ 
35 & 801.546 &     &    &       &     &    &       &           & Y &      \\
\it{36} & \it{802.654} & & \it{22} &  & \it{011} & \it{21} & \it{13 9} & \it{13104.6565} & \it{N} & \it{3000} \\
37 & 892.122 &      &    &       &     &    &       &           &Y  &      \\ 
38 & 892.160 & 100 & 27 & 11 16 &001 & 26 &9  17 & 13034.8501 & N &2975\\ 
39 & 892.198 & 010 & 26 & 13 13 & 010 & 25 & 12 14 & 11398.7613& Y&  1500\\ 
40 & 892.228 & 010 & 25 & 25 0 & 010 & 24 & 24 1 & 14644.7869&Y&  3275\\ 
41 & 892.286 & 000 & 29 &12 18 & 000 & 28 & 11 17& 11087.9927&Y&1500\\ 
42 & 892.345 & 030 & 16 & 5 12 & 030 & 15 & 2 13 & 7690.7488&Y&  1500\\ 
43 & 892.507 &     &    &       &     &    &       &           & Y &      \\
44 & 892.578  & 001 & 26 & 23 3 & 001 & 25 & 22 4 & 16039.7119&Y&  3515\\ 
45 & 892.634 & 001 & 27 & 16 12 & 001 & 26 & 15 11 & 14531.6796&Y  &3250\\ 
46 & 892.724 & 100 & 26 & 19 8 & 100 & 25 & 18 7 & 14815.6577&Y&  3300\\ 
47 & 892.760  & 100 & 26 & 23 4 & 100 & 25 & 22 3 & 16021.8356&Y&  3500\\ 
48 & 892.805 & 000 & 25 & 23 2 & 000 & 24 & 22 3 & 12049.8071&Y&  2800\\ 
49 & 892.838 & 001 & 26 & 20 6 & 001 & 25 & 19 7 & 15161.0169&Y&  3350\\ 
\it{50} & \it{892.871} & & \it{26} & & & \it{25} & & \it{18864.6094} &\it{N}& \it{4000} \\  
51 & 892.891 & 010 & 25 & 15 10 & 010 & 24 & 14 11 & 11478.5602 & Y &1500 \\
52 & 893.937 & 000 & 25 & 19 6 & 000 & 24 & 18 7 & 10831.4717&Y  &1500\\ 
53 & 894.184 & 020 & 31 & 0 31 & 010 & 31 & 1 30 & 11439.9008  & Y & 1500 \\
54 & 894.214 & 010 & 27 & 12 15 & 010 & 26 & 11 16 & 11673.7867&Y&  2750\\ 
\it{55} & \it{894.283} &  & \it{26} &  &  & \it{25} &  & \it{18588.1034}&\it{N}&  \it{3950}\\ 
56 & 894.416 &     &    &       &     &    &       &           & N &      \\
\end{tabular}
\end{table*}
\setcounter{table}{1}
\begin{table*}
\centering
\caption{{\it --continued                                                 }}
\begin{tabular}{|ccccccccccc|}
\hline
 & Frequency (cm$^{-1}$) & $\nu_1 \nu_2 \nu_3$ (u) & J (u) & K$_a$ K$_c$ (u) & $\nu_1 \nu_2 \nu_3$ (l) & J (l) & K$_a$ K$_c$ (l) & E(l) & Sunspot & T$_{max}$ (K) \\
\hline
57 & 894.542 & 010 & 19 & 7 13& 010 & 18 & 4 14& 6095.5140&Y&  1500\\ 
58 & 894.637 & 000 & 20 & 8 13 & 000 & 19 & 5 14 & 5052.6689&Y  &1500\\ 
{\it 59} & \it{894.697} &  & \it{29} & & & \it{28} & & \it{19917.3223} & \it{N} & \it{4000} \\
60 & 894.746 & 010 & 23 & 7 16 & 010 & 22 & 6 17 & 8181.3570 &Y &1500\\ 
61 &894.806  & 020 & 24 &15 10  & 020 & 23 &14 9  &12832.5041&Y  &2950\\ 
62 & 894.920 & 100 & 18 & 4 15 & 100 & 17 &1  16 & 6885.4857 & Y & 1500 \\ 
63 & 895.026 &     &    &       &     &    &       &           & N &      \\
64 & 895.070 &     &    &       &     &    &       &           & Y &      \\
{\it 65} &  \it{895.094} & & \it{22} & & & \it{21} & & \it{15222.2697} & \it{N} & \it{3375} \\
66 & 895.154 & 100 & 20 & 8 13& 100 & 19 & 5 14 & 8618.8825 &Y&1500 \\ 
67 & 895.203  & 001 & 20 & 5 15 & 001 & 19 & 4 16 & 8286.0153 &Y& 1500\\ 
68 & 895.252 & 000 & 25 & 22 3 & 000 & 24 & 21 4 & 11756.8965&Y&  2750\\
69 & 895.335 &     &    &       &     &    &       &           & N &      \\
70 & 895.376 &     &    &       &     &    &       &           & N &      \\
71 & 895.423$^1$ & 100    &13    &9 5       & 100   &12   & 6  6   & 6051.2728          & Y &  3375    \\ 
\hline
\multicolumn{11}{l}{$^1$ assigned by Zobov et al. (private communications) in the sunspots.} \\
\end{tabular}
\end{table*}
\subsection{Temperature sensitive water lines: a theoretical study} 

Once the water lines have been assigned, we attempt to determine the sensitivity
of each pure rotational transition to small (25~K) changes in temperature. This can be
determined by simply estimating the {\it relative} intensity of the lines as
a function of temperature by the use of modified Boltzmann equation:
\[ I_{rel} \sim 10^{-15} \frac{g_{lower}}{U} exp(-\frac{E_{lower}}{kT}) \] 
where $g_{lower}$ and $E_{lower}$ are respectively the statistical
weight and the energy level for the lower rotational state, and U is the
total partition function (which is a function of the temperature, T).
Note that our choice of 25K is rather arbitrary: in order to use 
pure rotational water lines as good tools for the determination of the
effective temperature of low mass stars, we require these lines to be sensitive to temperature changes below the temperature discrepancy found by other methods, usually at least 100K.
We calculated the partition function using the fitting formula
of Harris et al. (1998) (Equation 12) which is suitable for high temperatures. 
We have considered temperatures ranging from 1500~K to 4000~K in steps of 25~K.
Examples of intensities and gradients 
versus temperature are given
in Figure 5 for two pure rotational water transitions: while the weak water transition at 797.424 cm$^{-1}$ (line 5 from Table 2)
exhibits a fairly low and flat gradient, the water transition at 894.637 cm$^{-1}$ (line 58 from Table 2) is very
sensitive to temperature changes up to $\sim$ 2800 K but it becomes
less sensitive at higher temperatures. 
The temperature of the maximum
gradient (i.e the temperature at which the transition is most
sensitive to temperature changes) for all lines is also given in Table 2
(last column).
We find that about 10 transitions out of the 36 pure rotational ones identified in the BT2 linelist
are sensitive enough to small (25K) temperature changes that they could potentially be used in the determination
of an effective temperature scale for low mass stars. Clearly, a 25K sensitivity is 
a strict requirement. A 50K sensitivity would still make our methodology more accurate than others and would lead to the use of a larger number of lines: from the 36 pure rotational ones identified here {\bf we find about 14}.

The use of the most temperature sensitive water transitions for the
determination of an accurate effective temperature scale of 
low mass
stars becomes practical once a range
of calibration objects has been observed with suitable measurements of water
transitions. Our two dwarf stars only span around 300 K and have inadequate
S/N for these purposes.  Nevertheless, we attempt here to
demonstrate the methodology by computing the intensity and gradient of the {\it pure
rotational} transitions from Table 2 at 3500K.
Our computations are shown in Table
3 which includes the subset of rotational lines and identifiable in the
Sunspot atlas. From this table we see that some of the most sensitive lines,
at this particular temperature, are around 894.6 cm$^{-1}$. In Figure
6 we plot the ratio of the two stellar spectra for a small wavelength
region centred at 894.6 cm$^{-1}$ and compare it to the ratio of two
synthetic spectra computed at 3200 and 3500K.
The strength of the `absorption' for some of the lines in the divided spectra
is a clear indication that the transitions identified from Table 3
are indeed the most sensitive to temperature variations. 
It should be noted that
the strongest most sensitive water lines occur towards lower temperatures.
Observations of cooler objects should be within the reach of instruments such as EXES on
SOFIA. The method can be calibrated with measurements of the growing number
of cool low-mass eclipsing binaries. 

Finally, an effective temperature scale for low mass stars can only be quantitatively
derived if effects such as stellar rotation, internal {\bf and atmospheric} structure and equation of state
are taken into consideration when generating the synthetic spectra. 

\section{Conclusions}
In this paper we present the first high resolution MIR spectra of M
dwarfs. Very high resolution observations of pure rotational water vapour
transitions in the MIR may provide us with a new methodology in the
determination of the effective temperatures for low mass
stars. We have used the latest state-of-the-art water linelist (BT2, Barber et al. 2006) 
to identify and assign water transitions in these spectra. In total we assign 52 water 
lines out of 71 
likely water transitions; 36 are pure rotational lines. We have computed a theoretical sensitivity study to determine
how sensitive each pure rotational water transition is to small (25~K) changes in 
effective temperature and we find that at least 10 should be sensitive enough to be used
as temperature indicators.  

Due to the small wavelength region observed, the lack of a statistical
sample of objects spanning lower temperatures and the low S/N, 
it is impossible at this stage to determine the effective temperature
of our two M dwarfs using the methodology outlined in this work;
our technique is viable only provided we have a 
large enough number of sensitive rotational lines (hence a larger number
of wavelength regions in the 8-21 $\mu$m spectrum), 
as well as a large sample of objects spanning the
cool spectral subsequences. Both TEXES on Gemini 
and future generation telescopes such as the ELT have the potential to achieve
this.
\par
Beside the need of a larger sample of objects, ultimately, for a realistic determination of an
effective temperature scale,  
model atmospheres including the
BT2 water linelist should be used for future analysis.

\begin{table*}
    \caption{Relative intensity of pure rotational and 
assigned water lines at 3500K (Column 3), and Gradient at 3500K (Column 4).}
    \begin{tabular}{|cccc|}

 & Frequency (cm$^{-1}$) & Intensity (3500K) & Gradient (3500K) \\
\hline

 2 &     797.188&     5.97E-21 &     9.48E-26 \\
 5&     797.424&     4.19E-22 &     3.24E-25 \\
 6 &     797.551&     3.05E-21 &     6.36E-25 \\
 9 &     798.122&     6.68E-21 &     1.08E-25 \\
18 &     799.919&     7.33E-22 &     4.51E-25 \\
21 &     800.169&     1.23E-21 &     5.75E-25 \\
23 &     800.280&     7.24E-21 &     2.86E-25 \\
25 &     800.427&     1.14E-21 &     5.58E-25 \\
28 &     800.552&     1.28E-21 &     5.84E-25 \\
29 &     800.611&     4.67E-22 &     3.47E-25 \\
30 &     801.132&     1.07E-21 &     5.43E-25 \\
31 &     801.159&     6.08E-22 &     4.06E-25 \\
32 &     801.219&     5.56E-21 &     2.02E-25 \\
33 &     801.252&     5.98E-21 &     9.27E-26 \\
34 &     801.360&     3.46E-21 &     5.98E-25 \\
39 &     892.199&     1.43E-21 &     6.08E-25 \\
40 &     892.228&     3.78E-22 &     3.03E-25 \\
41 &     892.286&     1.63E-21 &     6.32E-25 \\
42 &     892.345&     6.58E-21 &     8.00E-26 \\
44 &     892.578&     2.13E-22 &     2.05E-25 \\
45 &     892.634&     3.96E-22 &     3.12E-25 \\
46 &     892.724&     3.52E-22 &     2.89E-25 \\
47 &     892.760&     2.14E-22 &     2.06E-25 \\
48 &     892.805&     1.10E-21 &     5.49E-25 \\
49 &     892.838&     3.05E-22 &     2.63E-25 \\
51 &     892.891&     1.39E-21 &     8.59E-26 \\
52 &     893.937&     1.81E-21 &     6.48E-25 \\
54 &     894.214&     1.28E-21 &     5.85E-25 \\
57 &     894.542&     1.27E-20 &     2.55E-24 \\
58 &     894.637&     1.95E-20 &     6.34E-24 \\
60 &     894.746&     5.38E-21 &     2.47E-25 \\
61 &     894.806&     2.61E-21 &     4.71E-25 \\
62 &     894.920&     9.17E-21 &     9.86E-25 \\
66 &     895.154&     4.50E-21 &     4.38E-25 \\
67 &     895.203&     5.16E-21 &     3.00E-25 \\
68 &     895.252&     1.24E-21 &     5.77E-25 \\

\end{tabular}
\end{table*}

\begin{figure*}
\includegraphics[width=160mm]{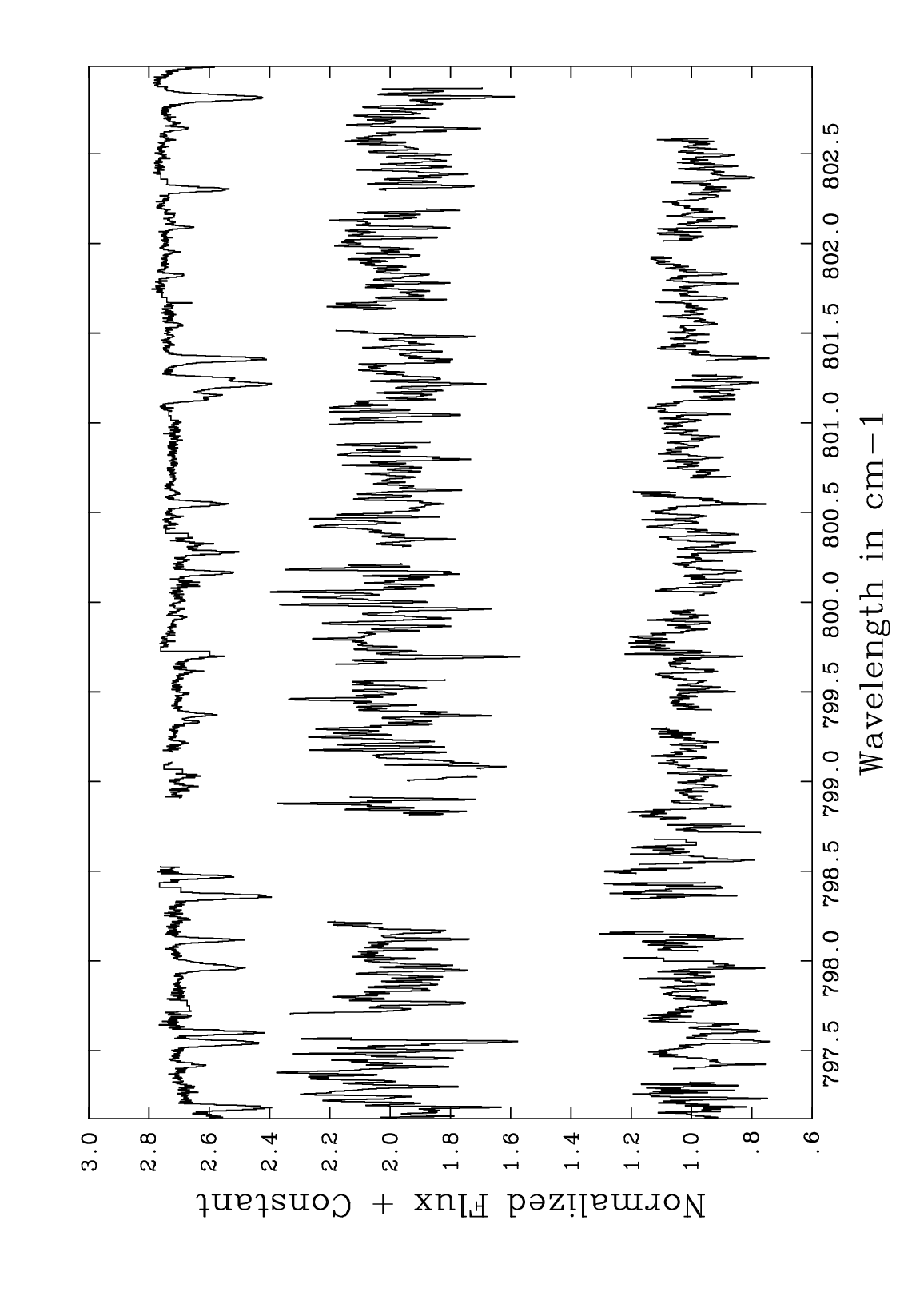}
\caption{Observed spectra centred at 11.2 $\mu$m. Top: BS587; Middle: GJ273; Bottom: GJ411. Spectra have been corrected
for stellar velocities (see text).}
\label{Figure1}
\end{figure*}

\begin{figure*}
\includegraphics[width=160mm]{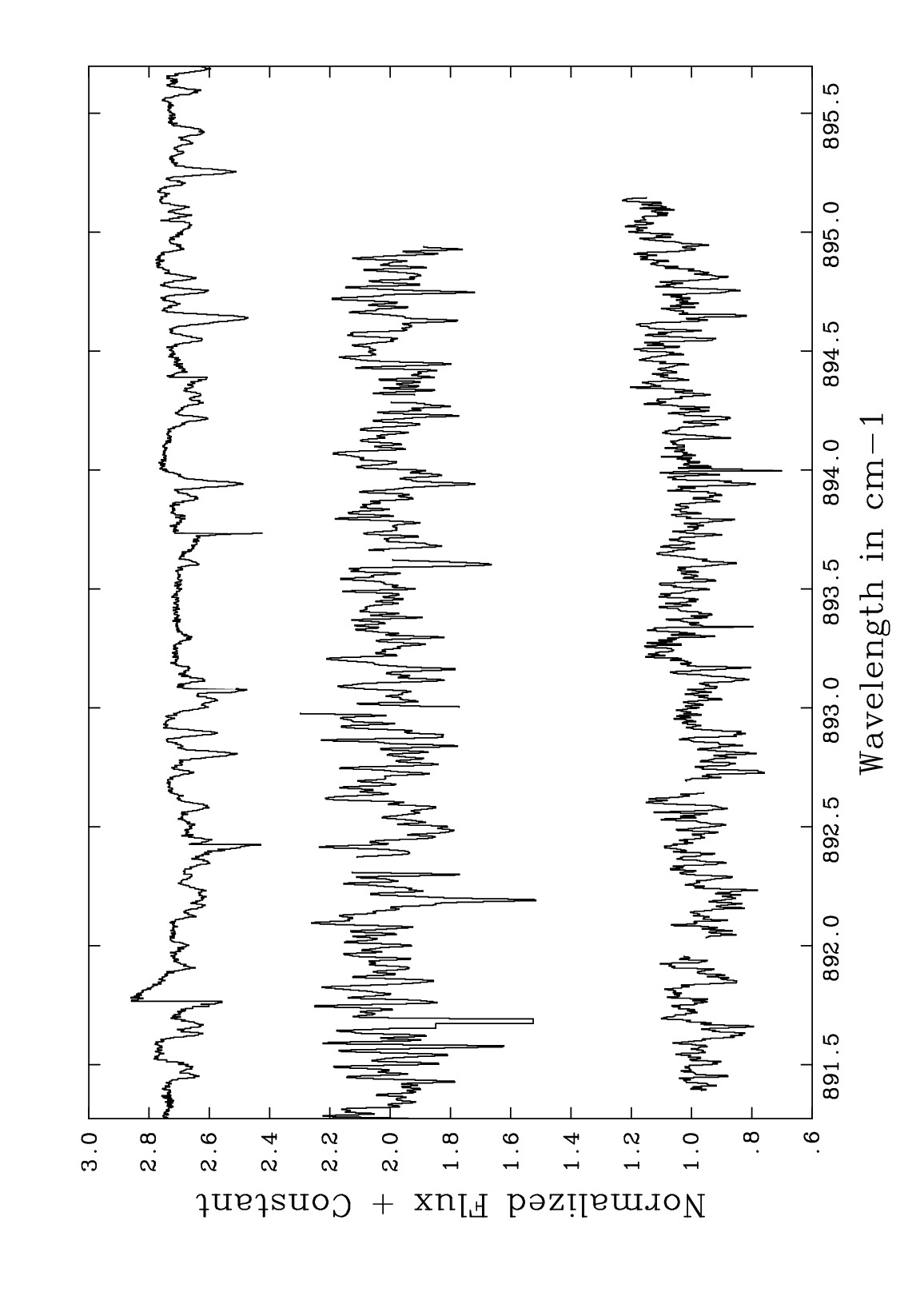}
\caption{As for figure 1 for the 12.5 $\mu$m region}

\label{Figure2}
\end{figure*}

\begin{figure*}
\vspace{-1cm}
\includegraphics[width=115mm, angle=-90]{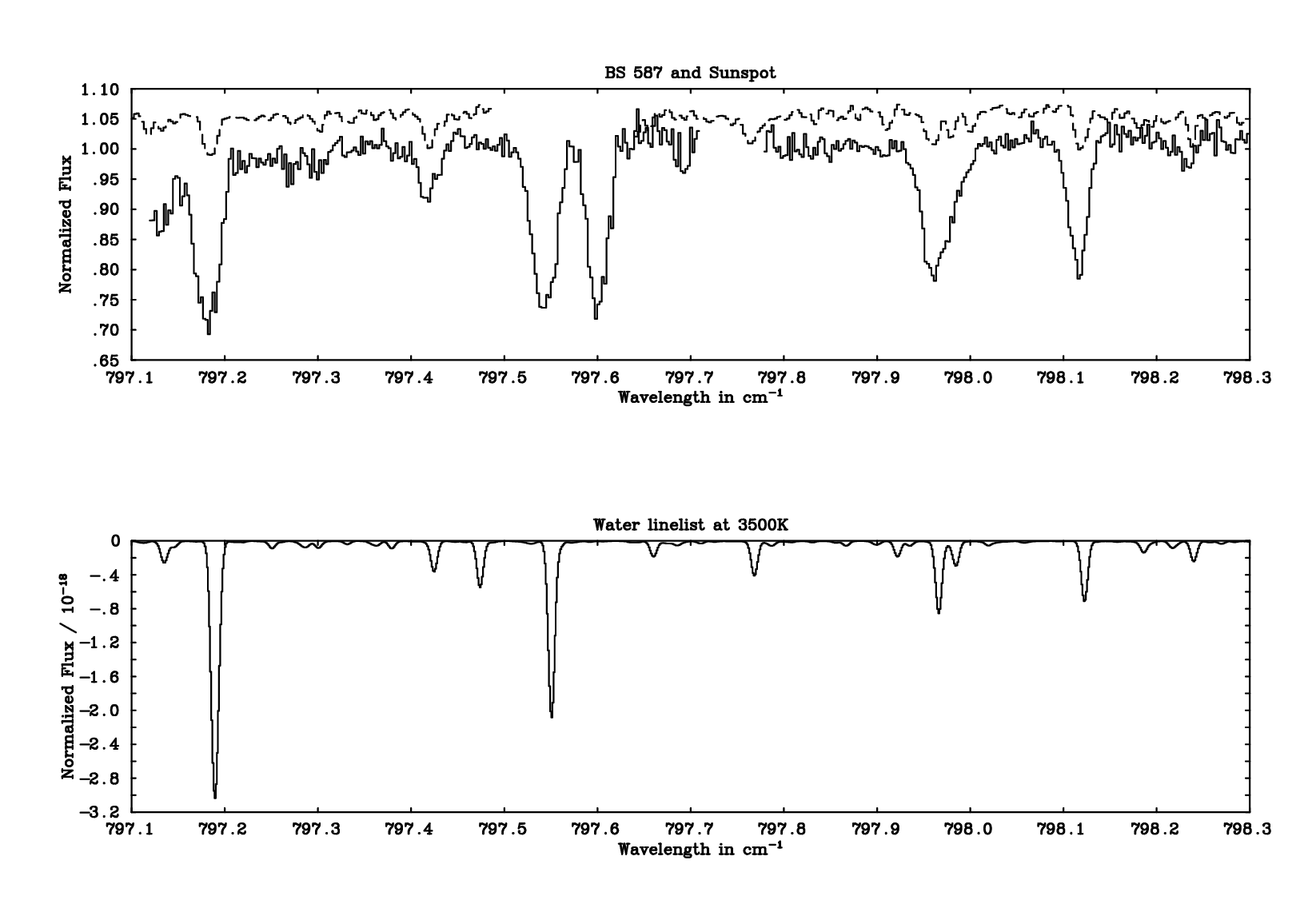}
\includegraphics[width=115mm, angle=-90]{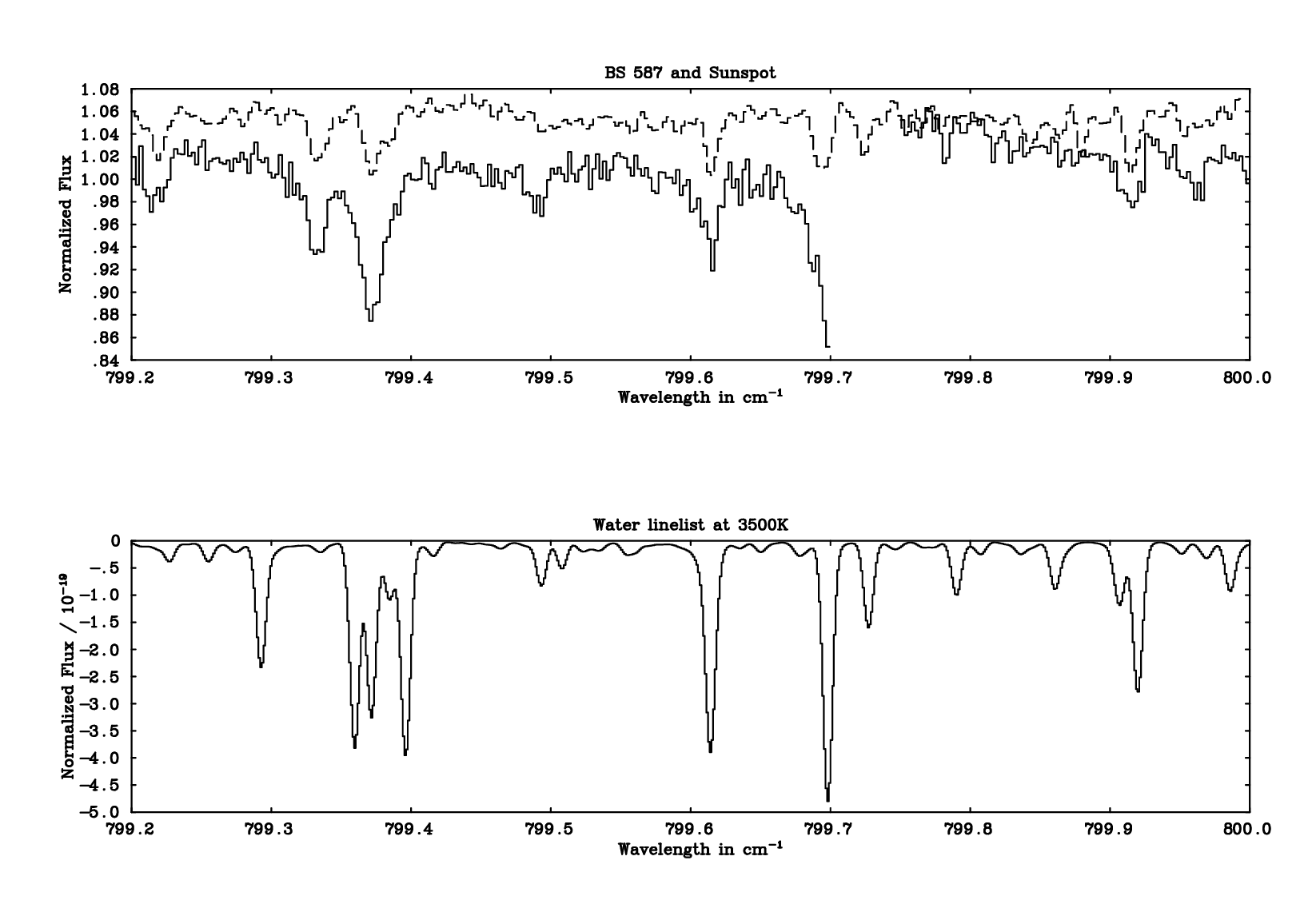}
\caption{Selected portions of spectra in the 11.2 $\mu$m region. Top: BS587 (Continuous) overplotted to
the Sunspot spectrum (dashed, flux offset by +0.05 for clarity); Bottom: synthetic BT2 linelist computed at 3500K. }
\label{Figure3}
\end{figure*}
 
\begin{figure*}
\vspace{-1cm}
\includegraphics[width=115mm, angle=-90]{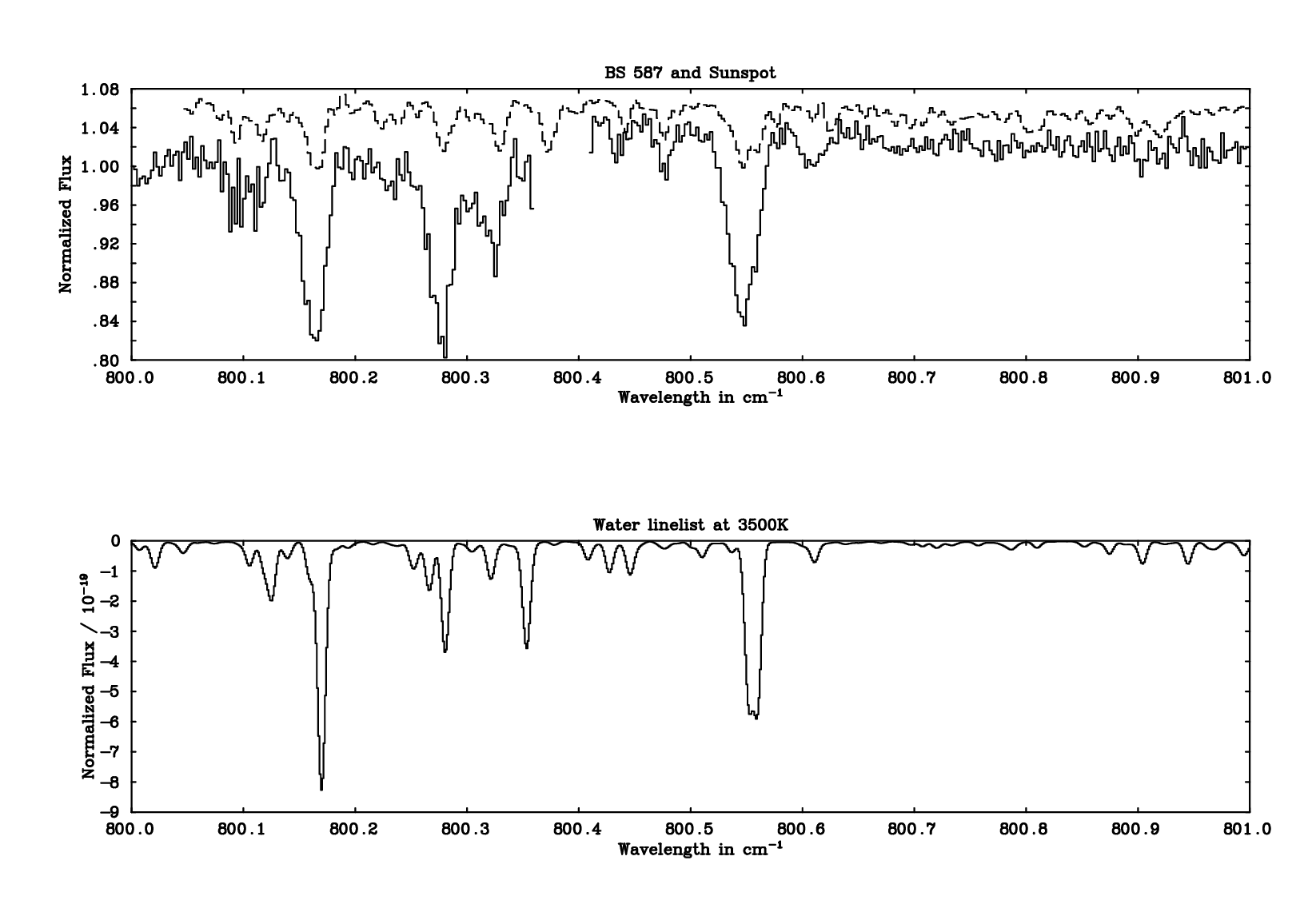}
\includegraphics[width=115mm, angle=-90]{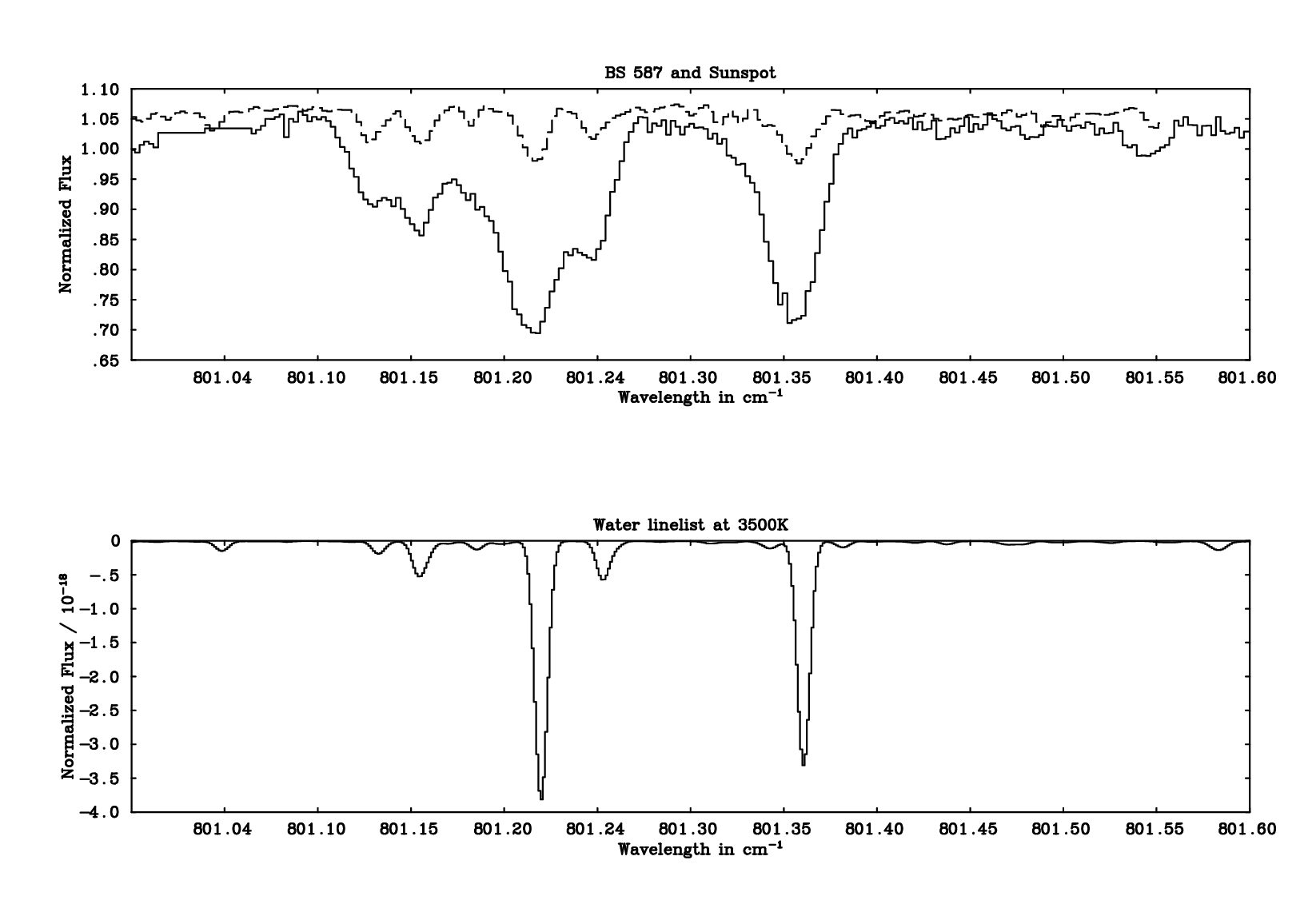}
\caption{As in Figure 3 but for the 12.5 $\mu$m region}
\label{Figure4}
\end{figure*}

\begin{figure*}
\vspace{-1cm}
\includegraphics[width=120mm, angle=-90]{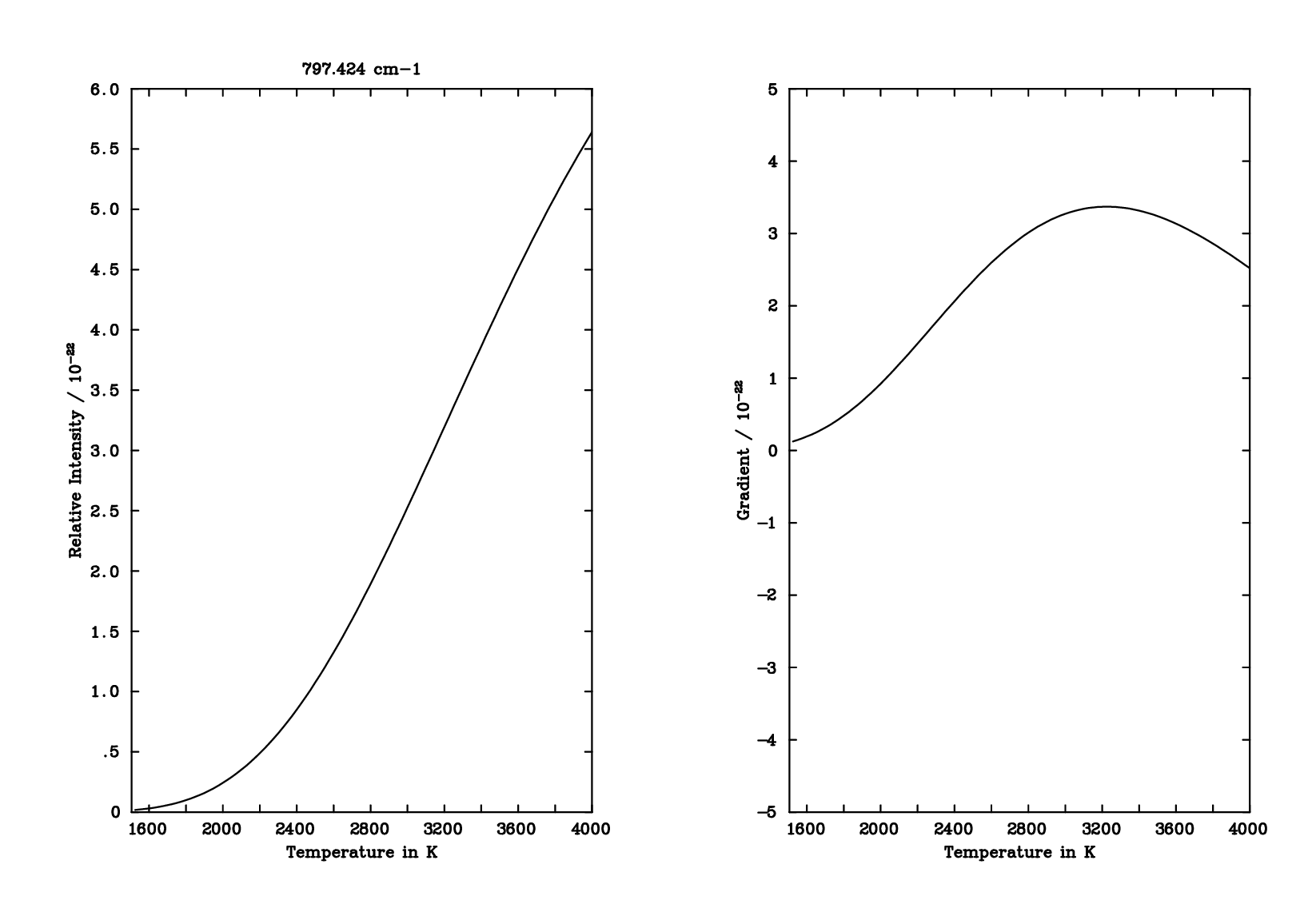}
\includegraphics[width=120mm, angle=-90]{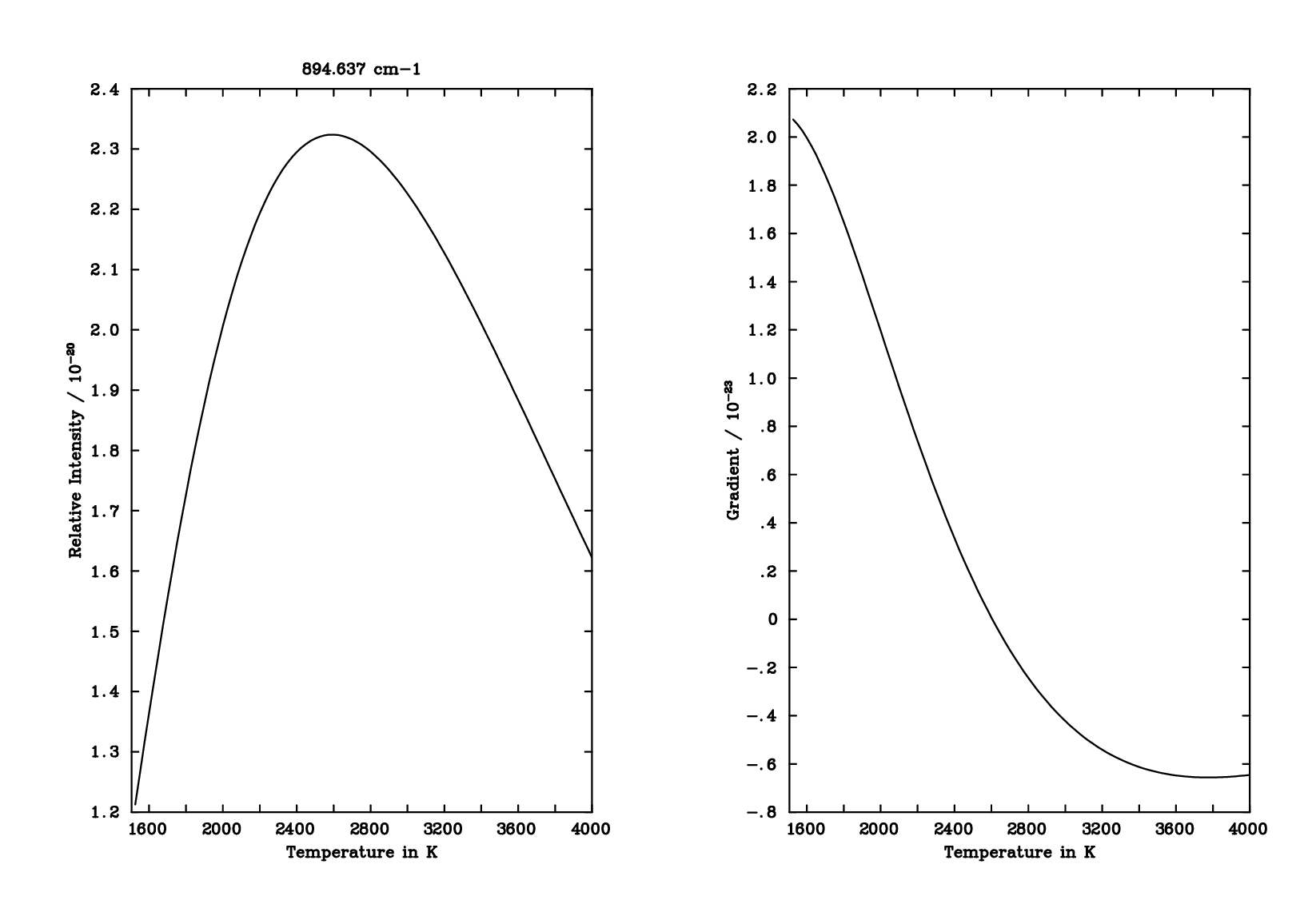}
\caption{Relative Intensity (e.g. Column 2 in Table 2) versus temperature (Left) and actual gradient 
versus temperature (Right) for a weak (Top) and a strong (Bottom) pure rotational water transitions from Table 3. The top gradient has been multiplied by a factor of 10$^{3}$ to fit within the same scale as the bottom one.}
\label{Figure5}
\end{figure*}

\begin{figure*}
\includegraphics[width=120mm,angle=-90]{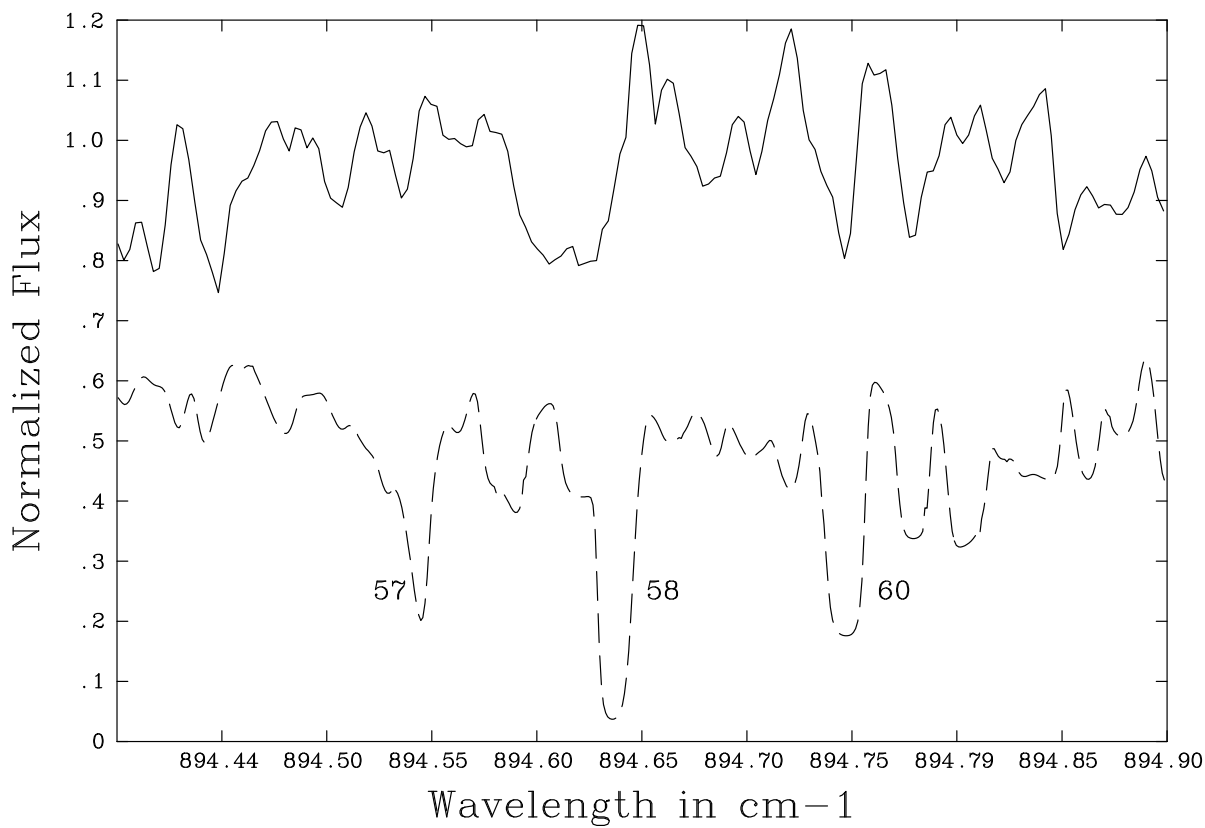}
\caption{Ratio of the GJ273 and  GJ411 spectra (Top) compared to a ratio of two synthetic spectra computed using the BT2 water linelist, at 3200 and 3500K.}
\label{Figure6}
\end{figure*}

\section{Acknowledgements}
Based on observations obtained at the Gemini Observatory, which is operated by the
Association of Universities for Research in Astronomy, Inc., under a cooperative agreement
with the NSF on behalf of the Gemini partnership: the National Science Foundation (United
States), the Science and Technology Facilities Council (United Kingdom), the
National Research Council (Canada), CONICYT (Chile), the Australian Research Council
(Australia), CNPq (Brazil) and SECYT (Argentina). 
Observations with TEXES (GN-2006B-Q-49) were supported by NSF grant AST-0607312.
SV acknowledges financial support from an individual PPARC Advanced Fellowship.  
MJR acknowledges NSF grant AST-0708074. We thank the referee for constructive comments which helped to improve the manuscript.


\end{document}